\documentclass[apj]{emulateapj}
\pdfoutput=1

\usepackage{graphicx}
\usepackage{subfigure}
\usepackage{amsmath}
\usepackage{amssymb}
\usepackage{xcolor}
\usepackage{url}
\usepackage{hyperref}
\usepackage{natbib}
\usepackage{euler}

\begin{document}

\title{Collisions of CO$_2$ Ice Grains in Planet Formation}

\author{Grzegorz Musiolik$^*$, Jens Teiser, Tim Jankowski, Gerhard Wurm}
\email{$^*$ E-Mail: gregor.musiolik@uni-due.de}

\address{Fakult{\"a}t f{\"u}r Physik, Universit{\"a}t Duisburg-Essen, Lotharstr. 1, 47048 Duisburg, Germany}

\begin{abstract}
In protoplanetary disks, CO$_2$ is solid ice beyond its snow line at $\sim 10 \rm AU$. Due to its high abundance, it contributes heavily to the collisional evolution in this region of the disk.
For the first time, we carried out laboratory collision experiments with CO$_2$ ice particles and a CO$_2$-covered wall at a temperature of 80 K. Collision velocities varied between 0 - 2.5 m/s. Particle sizes were on the order of $\sim$ 100 $\mu$m. We find a threshold velocity between the sticking and the bouncing regime at 0.04 m/s. Particles with greater velocities but below 1 m/s bounce off the wall. For yet greater velocities, fragmentation occurs. We give analytical models for the coefficients of restitution and fragmentation strength consistent with the experimental data. 
Set in context, our data show that CO$_2$ ice and silicate dust resemble each other in the collisional behavior. Compared to water ice the sticking velocity is an order of magnitude smaller. One immediate consequence as example is that water ice particles mantled by CO$_2$ ice lose any "sticking advantage." In this case, preferential planetesimal growth attributed to the sticking properties of water ice will be limited to the region between the H$_2$O ice line and the CO$_2$ ice line.  
\end{abstract}

\section{Introduction}

The role of silicate dust and water ice in planetesimal formation by collisional growth has been studied for many years \citep{Johansen2014,Blum2008,Dominik2007}. The role of CO$_2$ ice and CO ice has not been touched upon yet in the framework of collisional evolution, although carbon oxides are highly abundant. 
For comet Churyumov-Gerasimenko, \textit{Rosetta} recently found significant CO$_2$ gas fractions in the coma composition \citep{Haessig2015}. \citet{Pontoppidan2014} and \citet{Oeberg2011} estimate the number ratios of H$_2$O:CO$_2$:CO and give values of 100:38:31 in cloud core environments and 100:13:13 toward protostars.

In the general scheme of planetesimal formation, the ices of these volatiles play a role beyond their snow lines.
\citet{Alidib2014} suggest that the formation of Uranus and Neptune might be tight to the CO ice line. For TW Hya, observations by ALMA imply the existence of a CO ice line at $\sim$30 AU \citep{Qi2013}. 
Further inward though is the CO$_2$ ice line.
Its position can be calculated using the sublimation pressure curve and a disk model. We fitted the experimental sublimation pressure data from \cite{Bryson1974} with a modified sublimation pressure curve from \citet{Span1994}, leading to
\begin{equation}
\ln\left( \frac{p_{\rm sub}}{p_{\rm c}} \right) = \frac{T_{\rm c}}{T} \left[ k_1 \left( 1 - \frac{T}{T_{\rm c}} \right)^{l_{1}} + k_2 \left( 1 - \frac{T}{T_{\rm c}} \right)^{l_{2}}  \right]
\label{co2subpressure}
\end{equation}
with $T_{\rm c} = 216.592$ K, $p_{\rm c} = 0.51795$ MPa, $k_1 = -12.3081445$, $k_2 = -5.3061778$, $l_1 = 0.8472$, $l_2 = 3.47023$. 
Using the minimum mass solar nebula (MMSN) disk properties by \citet{Hayashi1985} and setting $p_{\rm sub} = p_{\rm disk}$, the CO$_2$ snow line can be calculated to be at 9.3 AU. With the same method, the H$_2$O snow line is calculated to be at 2 AU. Using different models and calculation methods, the CO$_2$ and H$_2$O snow lines might shift, but, in general, the values are consistent with those from other authors like \citet{Oeberg2011} for example.

Essentially no data exist to date on the collisional behavior of CO$_2$ ice grains. We note that there are investigations related to CO$_2$ snow. These have applications on Earth in mind, e.g. \citet{Lin2014} explain an agglomeration mechanism of 100 $\mu$m sized CO$_2$ snow inside a tube and describe its dependency on jet flows with vortices.
This cannot be applied to astrophysical environments directly, though. Quantitative experimental studies similar to those made for planetesimal formation with silicates \citep{Blum2008,Wurm2005,Teiser2011,Guettler2010} or H$_2$O ice \citep{Aumatell2011, Aumatell2014, Gundlach2015} are missing so far. 

In the context of planetesimal (or comet) formation, it is important to know threshold velocities between sticking and bouncing as a function of grain size for all relevant condensed species. It is also important to have an estimate of how much energy is dissipated in a bouncing collision. Last but not least, the onset and strength of fragmentation are also
fundamental properties to know. \citet{wettlaufer2010} e.g. considers collisional fusion of ice, where dissipation might lead to local heating and sintering, which might be important at higher temperatures. \citet{Wurm2001} consider reaccretion of particles after a collision by gas drag. Currently, this is also detailed by \citet{jankowski2015} and the rebound speed would be important then.

\begin{figure*}[htb!]
	\centering
	\includegraphics[width=\textwidth]{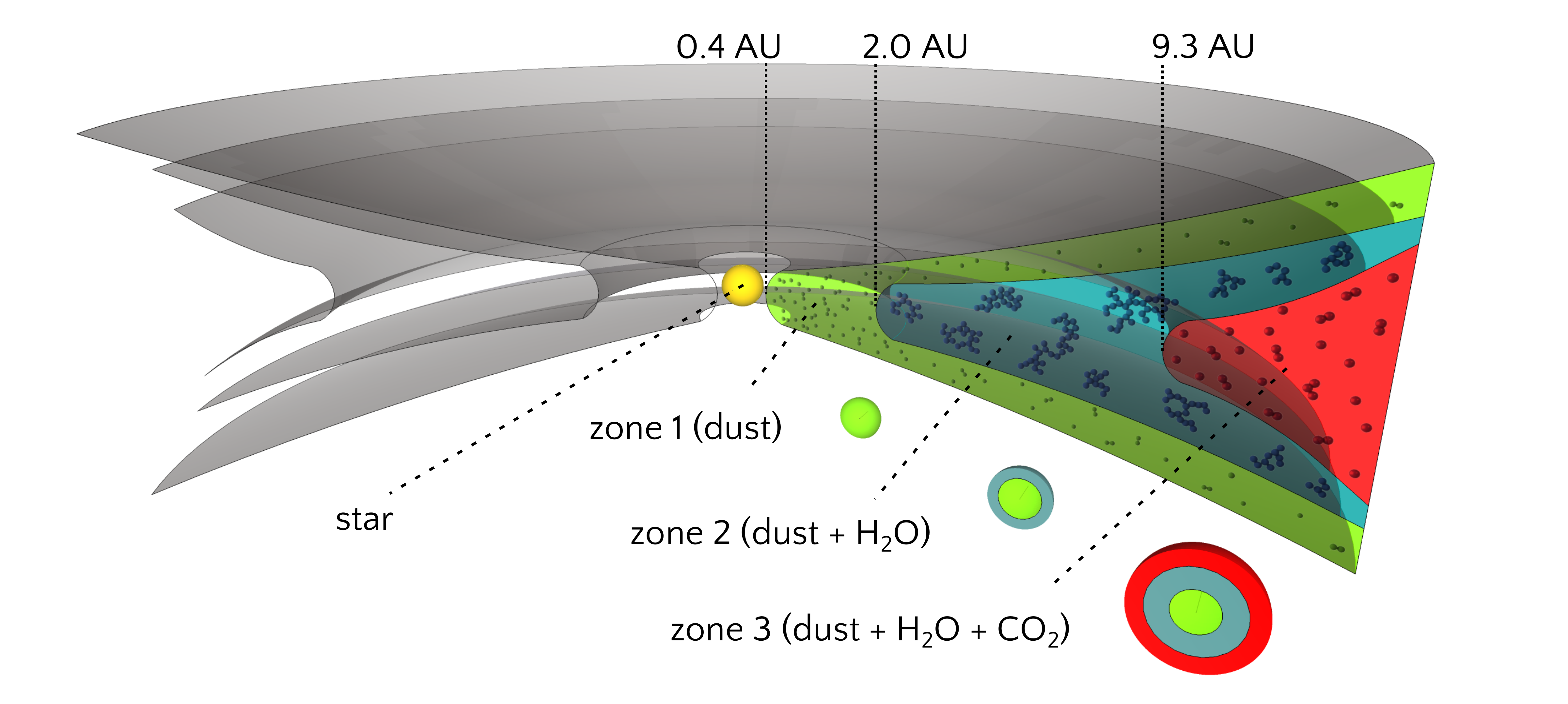}
	\caption{\label{ppde}
		Initial setting of a protoplanetary disk. Whilst in the first zone only silicate particles exist, in the second zone beyond the H$_2$O snowline the silicate particles might be covered by water ice. The CO$_2$ snow line indicates the beginning of the third zone, where silicate particles are coated with two layers: one inner H$_2$O layer and an outer CO$_2$ layer. Note that for better visualization the coatings are solid and colored, green: silicates; blue: H$_2$O; red: CO$_2$. While within the first, silicate-dominated zone as well as in the third, CO$_2$-dominated zone particles cannot grow quickly to larger sizes, H$_2$O coated particles in the second zone might grow to larger sizes more efficiently.}
\end{figure*}

It should be obvious, but we would like to stress that CO$_2$ and $\rm H_2O$ are quite different. Besides the general common feature of volatility that kind of labels both solid phases as ice, it is $a$  $priori$ not possible to equal the collisional behavior of CO$_2$- and $\rm H_2O$-ice without considering the microphysics of collisional particle interaction. In contrast to $\rm H_2O$, CO$_2$ molecules do not have any electrical dipole moment \citep{Morrison1979}. This is important for cohesion and collisional sticking. As shown in this paper, the term ice is not a synonym for sticky.  
Therefore, a possible zoned setting of a protoplanetary disk might look like fig. \ref{ppde}. The idea underlying this picture is an onion-like grain structure. The most refractory material is in the center; the most volatile ice is on the surface. This picture can be found throughout the literature \citep{Sirono2013, Fayolle2014}. As shown later, growth of particles might preferably happen in zone 2, where H$_2$O ice dominates the collisional behavior.

In a complex scenario of sublimation and recondensation, especially in the vicinity of the different ice lines, the picture of layered grains might be somewhat simplified. However, it will readily show the
potentially wide implications of the CO$_2$ collisional behavior. Mostly moderated by surface forces, the material of the outermost layer counts at least for major species like CO$_2$ that can can build sufficient mantle thickness. 

The composition of the mantle determines the range of velocities and particle masses / sizes for which
hit-and-stick collisions occur as grains collide \citep{Geretshauser2011}. For silicates in the warm region (zone 1 in fig. \ref{ppde}), growth via collisions easily works to millimeter size. However, from this size on, particles tend to bounce off each other rather than stick together. This is known as the bouncing barrier \citep{Zsom2010} and has also been shown to be a robust finding in experiments \citep{Kelling2014}. If large grains are introduced by some way collisional growth can still proceed \citep{Windmark2012, Wurm2005, Teiser2009, Deckers2014}. This is based on collisions with fragmentation where part of a smaller body is added to the larger one.

In the cooler regions of the disk (zone 2 in fig. \ref{ppde}) water ice is the next solid to dominate the collisional outcome. Due to the increased surface energy of the water ice with its polar molecules, bindings are much stronger and particles can grow larger. This has been seen in numerical simulations \citep{Dominik1997, Wada2009} and laboratory experiments \citep{Gundlach2015, Aumatell2014}. \citet{Okuzumi2012} even considered this to lead to very large planetesimal size porous ice aggregates. Follow up work by \citet{Kataoka2013} also supports this view. 
In total, zone 2 seems very favorable for collisional planetesimal-size formation due to the properties of water ice. 

We note that other concepts of planetesimal formation are also intensively studied currently. To span the gap from sub-meter particles to planetesimal  gravito-turbulent concentration, concentration in baroclinic instabilities or streaming instabiliies are considered to concentrate solids within the disk to a point of gravitational collapse \citep{Chiang2010, Dittrich2013, Johansen2014}. 
Both kinds of formation mechanisms -- collisional growth and gravitational instabilities -- might also go hand in hand. More efficient sticking might provide the necessary seeds of certain sizes for instabilities more easily \citep{Johansen2014}

To date, also partly due to the increased mass reservoir of solids, giant-planet cores are most often supposed to be formed beyond the water snow line and formation models are typically set in this regime \citep{Pollack1996, Helled2014}. Close-in giant planets are thought to be migrated there afterward \citep{Baruteau2014, Kley2012, Dittkrist2014}.
In any case, the difference between zone 1 and zone 2 is evident with zone 2 being the one supposed
to favor rapid formation of larger bodies.

Not considered so far is how far zone 2 extends. Further outward in a disk, other ice lines exist. As
outlined above, CO$_2$ is very abundant. We consider it to be the next major ice line.
In the picture of mantled grains, colliding grains beyond this line will no longer be dominated by water-ice properties. It is therefore important to know how the picture changes for CO$_2$ particles.

In this paper, we quantify collisional outcomes between $\sim$ 100 $\mu$m sized CO$_2$ particles and a CO$_2$ covered wall at a temperature of 80 K and ambient pressure of 0.5 mbar. The collision velocities reach from 0 to 2.5 m/s. We determine the impact velocity thresholds between sticking, bouncing, and fragmentation. Furthermore, we study the fragment size distribution when critical fragmentation occurs. We present an analytical model for fragmentation and restitution that describes the measured collisional behavior, which we finally put in context of planetesimal formation.

\section{Experimental setup}
\begin{figure*}[htb!]
\centering
\includegraphics[width=\textwidth]{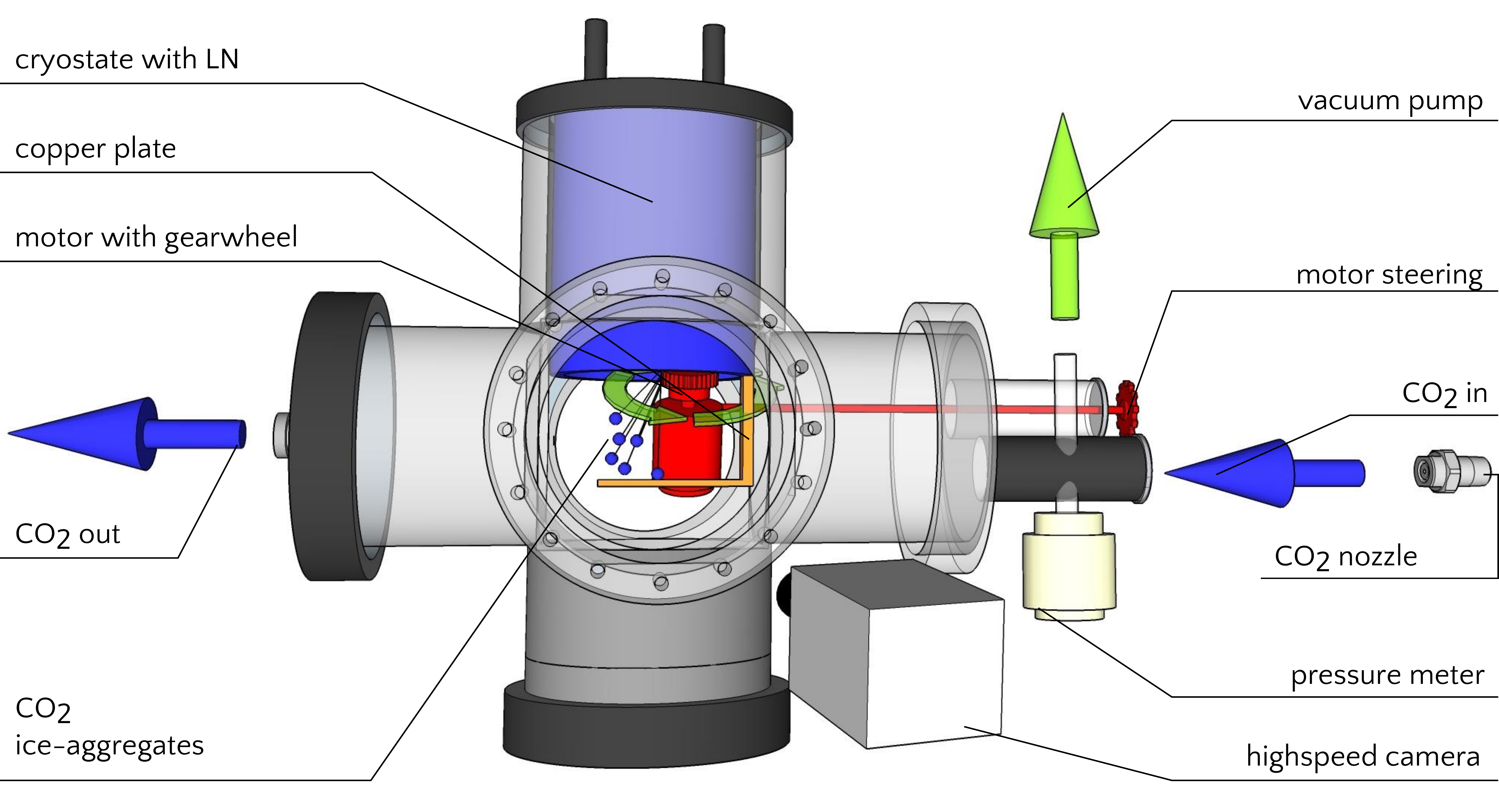}
\caption{\label{setup}
Experimental setup. Within a vacuum chamber, a liquid-nitrogen-cooled cryostat cools the top flange. The chamber is flooded with gaseous CO$_2$ which is deposited at the cryostat and a thermally coupled copper plate below the cryostate, forming an approximately 2 mm thick layer of solid CO$_2$. Using a cogwheel placed on a motor, ice aggregates can be beveled off the cryostat and the
particles collide with the CO$_2$ layer at the copper plate.}
\end{figure*}

The setup of the experiment is shown in fig. \ref{setup}.
The experiment consists of a vacuum chamber, a liquid nitrogen cooled cryostat with a connected copper plate 15 cm below, a CO$_2$ injecting nozzle, and a motor with a cogwheel. 

\begin{figure}[h]
	\includegraphics[width=\columnwidth]{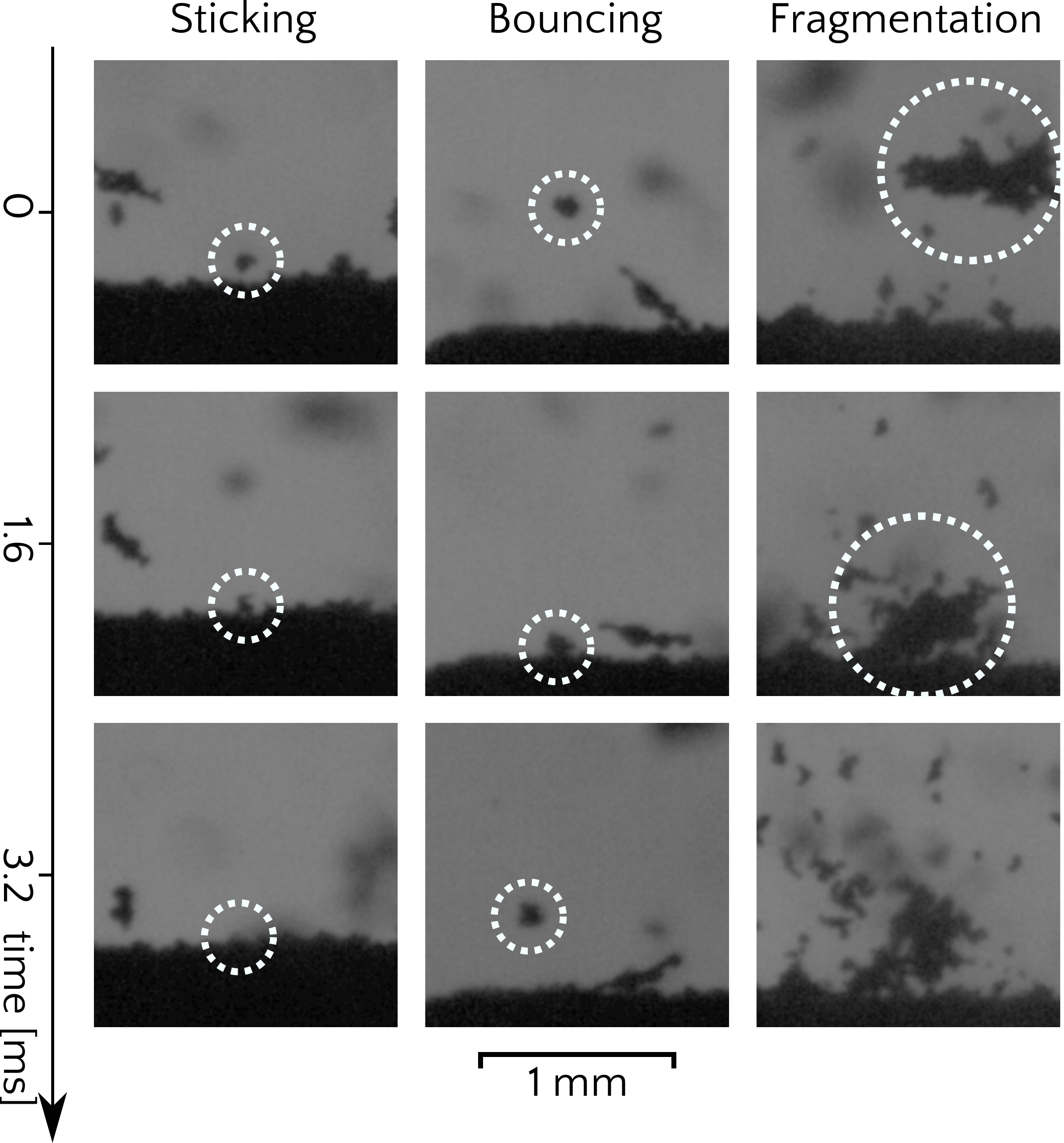}
	\caption{\label{camera} Sample images from the data set. CO$_2$ particles coming from the top are impinging the CO$_2$ ice layer at the bottom. Aggregates with surface equivalent radii greater than 500 $\mu$m were not studied. The examples show time sequences of the typical collisional outcomes observed: sticking, bouncing, fragmentation}
\end{figure}

As first step, the vacuum chamber is flooded with CO$_2$ gas. Since CO$_2$ gas is deposited on surfaces below 195 K for atmospheric pressures, a 2 mm thick, solid CO$_2$ ice layer grows on the walls of the cryostate and on the copper plate within several minutes.
In a second step, the vacuum chamber is sealed and evacuated to an ambient pressure of 0.5 mbar. Using the motor with the gearwheel, which can be slided and rotated along the chamber from the outside, CO$_2$ ice can then be beveled off the cryostate and collide at speeds below 2.5 m/s with the CO$_2$ layer on the copper plane. Low-velocity collisions at $\sim$0.05 m/s can be observed due to rebound and gas-grain coupling of the aggregates. The collisions are imaged using a high-speed camera with a microscope lens recording at 1250 frames per second. The spatial resolution is limited to 10 $\mu$m.
Sample images from the data set are shown in fig. \ref{camera}.

In total, we performed around 20 experiments and evaluated 96 collision events therein.

\section{Results and discussion}
\begin{figure}[h]
	\includegraphics[width=\columnwidth]{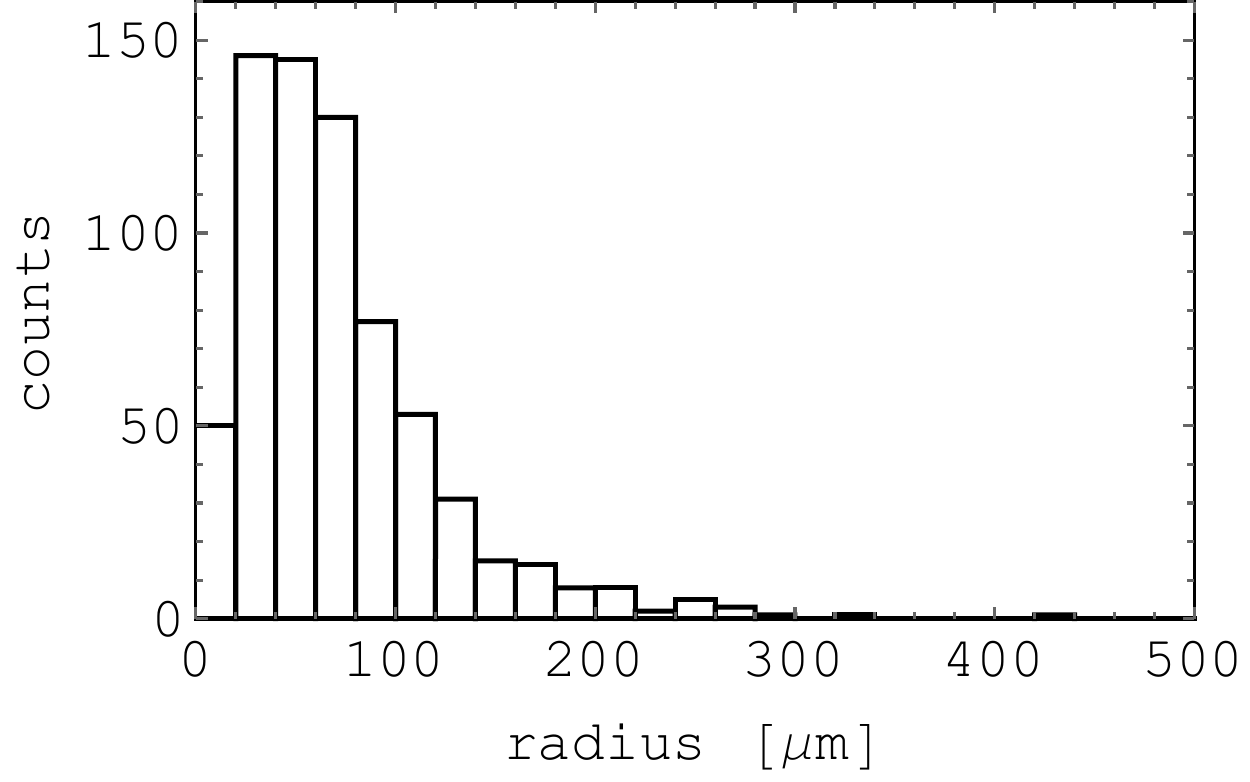}
	\caption{\label{size} Size distribution of the CO$_2$ particles. 690 particles were analyzed. The average size is $71\, \rm \mu m$. The bin size of 20 $\mu$m is chosen with respect to the spatial resolution of the optical system}
\end{figure}

For further analysis, we consider the particles to be spheres with a density of 1560 kg/m$^3$ \citep{Mazzoldi2008}. We take the radius of a sphere with the same cross section as the observed particle. Fig. \ref{size} shows the distribution of the particle radii used for this study.

Irregular grains might behave differently in individual collisions compared to spherical particles. \citet{Blum2000} and \citet{Poppe2000} suggest irregular grains of same size to be slightly stickier. However, it is mostly the size that determines the threshold velocity for sticking. We therefore consider our collisions to be representative of collisions of solid $\sim$100 $\mu$m CO$_2$ grains. This holds for the sticking and bouncing collisions. It should be noted that we could not unambiguously determine if the grains are monolithic or aggregates and the scaling to smaller sizes should only be considered as an order of magnitude estimate for these first experiments. For fragmentation at higher speed, the outcome is influenced by the internal binding. If the larger grains (see fig. \ref{camera} right) are aggregates bound by surface forces, they fragment easily. If all connections are solid, they might not. Due to the formation mechanism of scraping grains from a solid surface, we do
not know this {\it a priori}. However, the way the particles fragment into a large number of smaller units at low velocities indicates an aggregate structure. Comparing energies below supports this view that we produce grains of $\sim$100 $\mu$m in diameter and aggregates thereof.

The collision behavior of dust depends mainly on the initial collision velocities and the masses of the grains and can be divided into several types \citep{Guettler2010}. 80$\%$ of all particles are in the size range between 40 and 120 $\mu$m. The average size of the particles is 71 $\mu$m. In terms of mass of individual grains and therefore energy ($R^3$) this is
a variation in particle energy by a factor of about $120^3/40^3 = 27$. The collision velocity varies between 0.05 m/s and 2.5 m/s, which in terms of energy ($v^2$) is a factor 2500 which is much larger than the energy variation by mass. Also, fig. \ref{CORsize} shows an extreme division for the coefficient of restitution for particles smaller and larger than 150 $\mu$m that show no difference. We therefore consider the velocity dependence of the coefficient of restitution to be representative of the average-size particles. 

The CO$_2$ particles show the same well-known qualitative behavior as silicate particles. At low velocities, the CO$_2$ particles "hit $\&$ stick." If the speed is greater than a sticking velocity $v_\text{stick}$, the aggregates bounce off the surface in inelastic collisions. For these collisions, neither the surface is eroded nor do the particles fragment. Only at a greater velocity than $v_\text{frag}$, does fragmentation start. In our experiment, we were not able to see mass transfer during fragmenting collisions.

\begin{figure}[h]
	\includegraphics[width=\columnwidth]{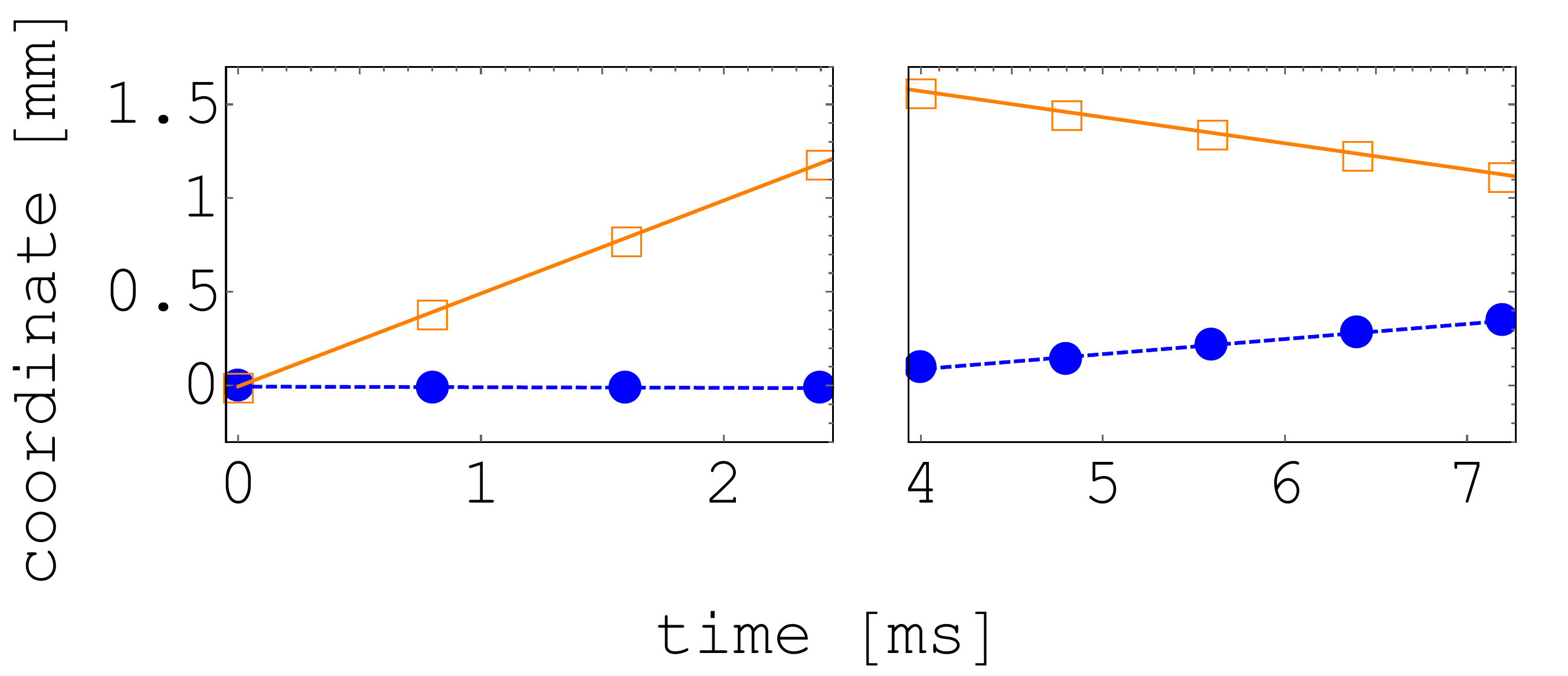}
	\includegraphics[width=\columnwidth]{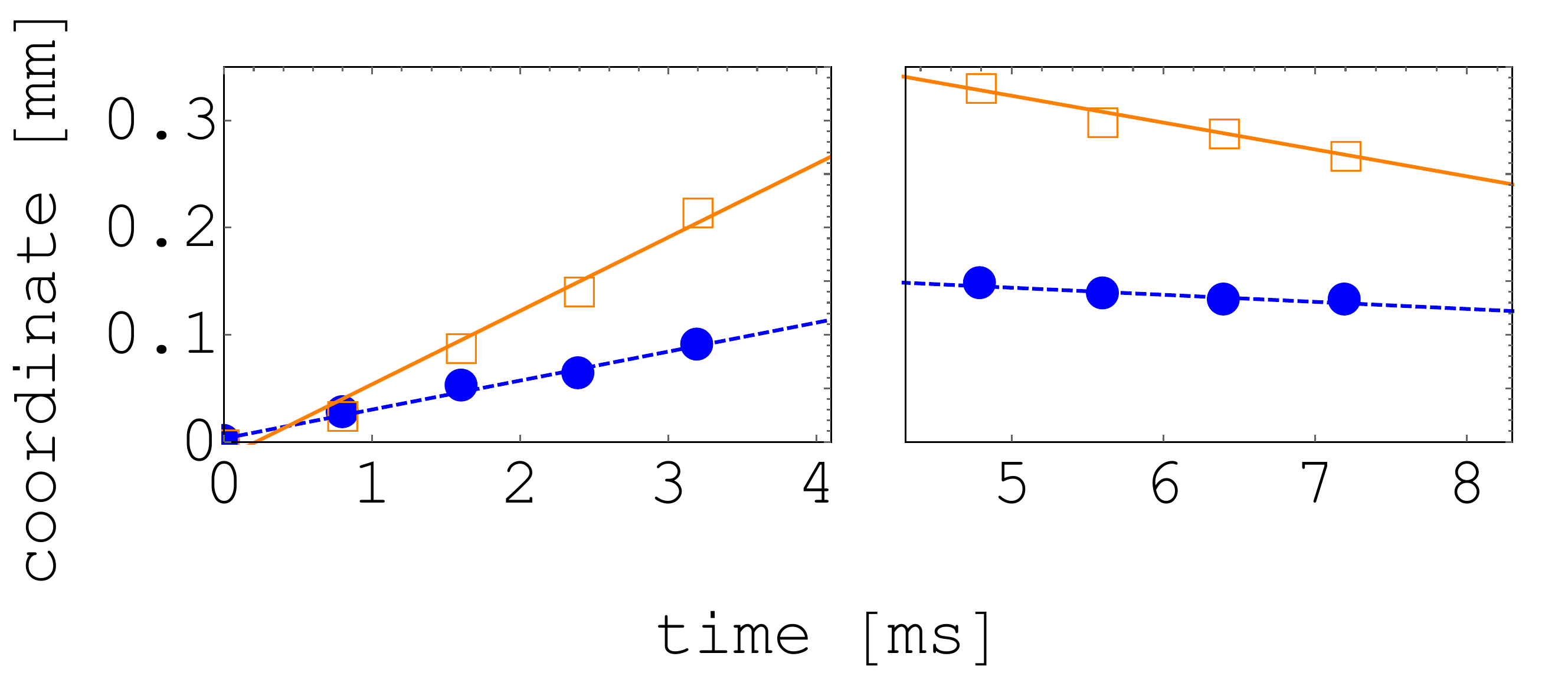}
	\caption{\label{linear} Two dimensional motion of a bouncing particle with overplotted linear fit functions (circles for camera x-coordinate, squares for camera y-coordinate) before the collision (left) and after the collision (right).  On top an example for higher velocities at about 1 m/s is shown. Below we give another example for velocities around 0.05 m/s.
	}
\end{figure}

The motion of a particle can be approximated as linear in a time frame of $\pm 5$ ms around the collision, like fig. \ref{linear} shows exemplarily for two bouncing particles.
By means of linear fits, we were able to determine the velocities of the particle before and after a collision $v_\text{i,x}$, $v_\text{i,y}$ and $v_\text{o,x}$, $v_\text{o,y}$ along the $x$ (horizontal) and $y$ (vertical) coordinates, respectively. We calculate the coefficient of restitution $\epsilon$, which we define as the ratio 
\begin{equation}
\epsilon(v_\text{o},v_\text{i}) = \frac{v_\text{o}}{v_\text{i}} = \left. \sqrt{v_\text{o,x}^2+v_\text{o,y}^2} \right/ \sqrt{v_\text{i,x}^2+v_\text{i,y}^2}.
\label{coefficientofrestitution}
\end{equation}
Moreover, we examined the fragmentation strength $\mu$, which is the ratio of the largest fragment's mass $m$ and the original aggregate mass $m_0$ after a fragmentation
\begin{equation}
	\mu(m,m_0)= \frac{m}{m_0}.
	\label{fragmentationstrength}
\end{equation}

\begin{figure}[h]
	\includegraphics[width=\columnwidth]{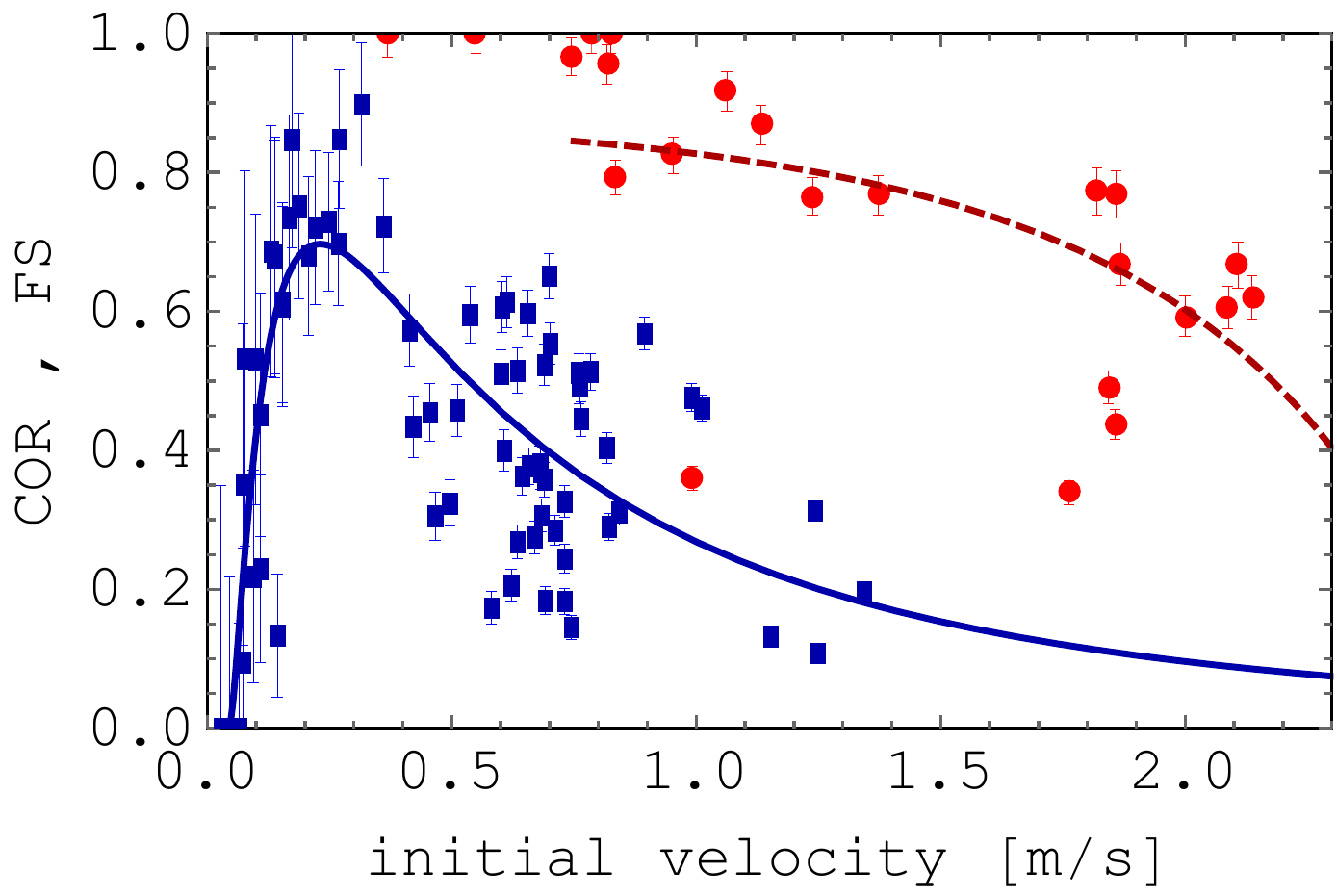}
	\caption{\label{transitionplot} Coefficient of restitution (COR; blue sqaures) and the fragmentation strength (FS; red dots) of 96 CO$_2$ collisions. The data are fitted using the models for restitution (solid blue line; eq. \eqref{restmodel}) and for fragmentation (dashed red line; eq. \eqref{fragmodel}). 
	}
\end{figure}

\begin{figure}[h]
	\includegraphics[width=\columnwidth]{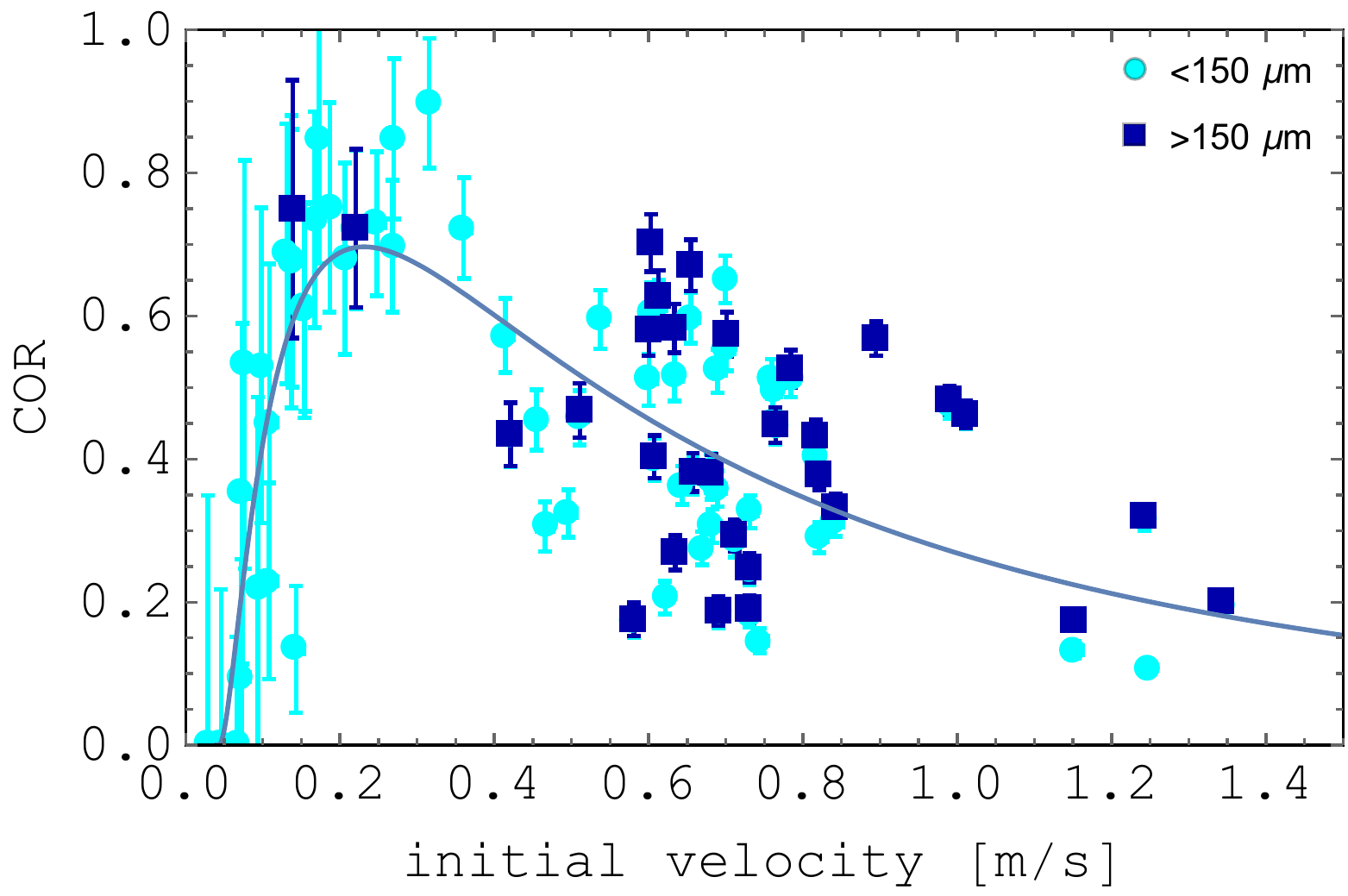}
	\caption{\label{CORsize} Coefficient of restitution for grains of different sizes. The separation is done in respect to the particles radii.
	}
\end{figure}

Fig. \ref{transitionplot} shows both quantities for all 96 observed collisions.
The coefficient of restitution was evaluated for all grain sizes from fig. \ref{size}. As shown in fig. \ref{CORsize} the larger grains collide at greater velocities due to diminished gas drag. Where both size ranges are present, the coefficient of restitution shows no difference within the studied total range. 

\begin{figure}[h]
	\includegraphics[width=\columnwidth]{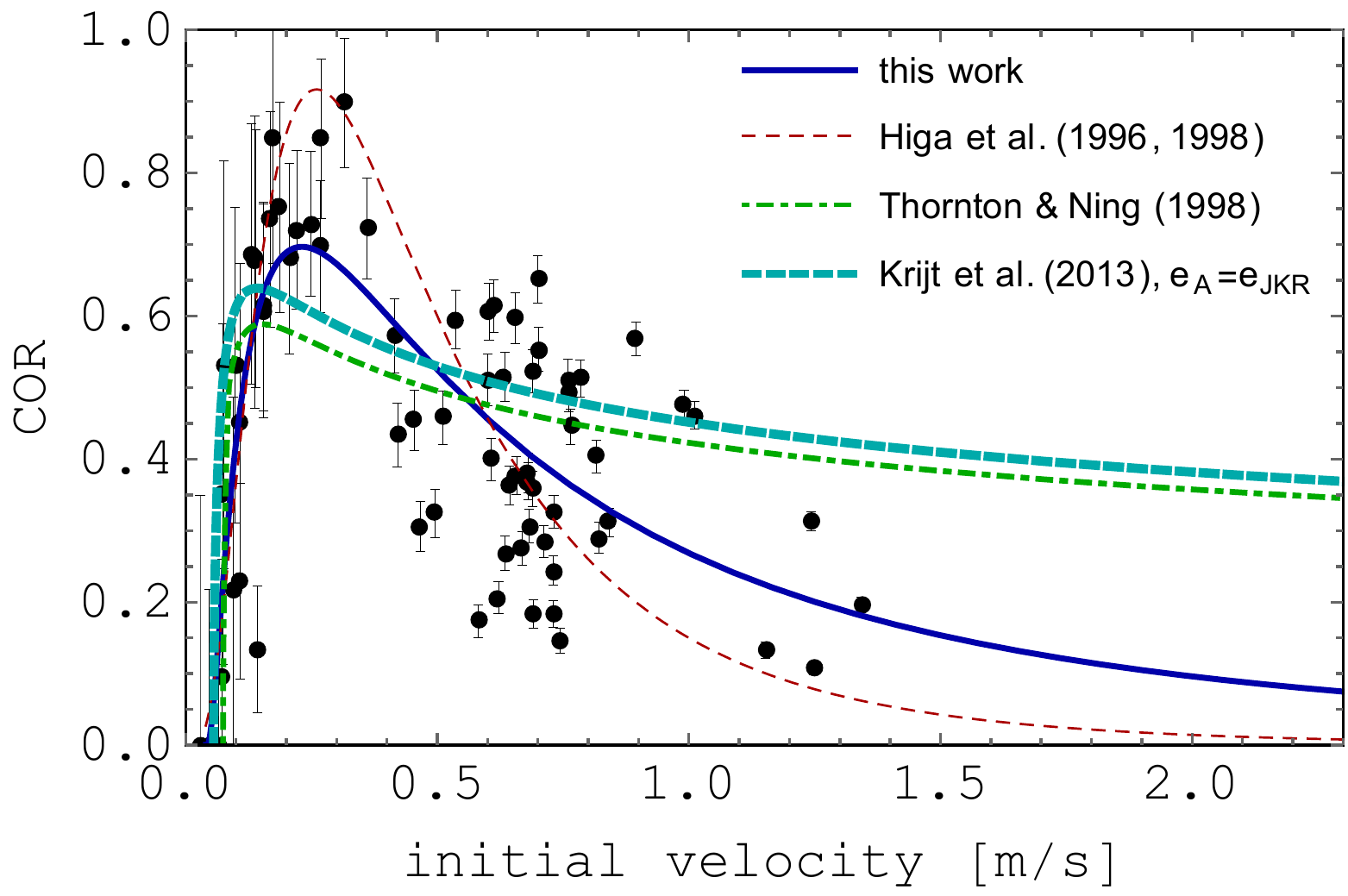}
	\caption{\label{cmod}Comparison of different models for the coefficient of restitution.
	}
\end{figure}

The coefficient of restitution peaks at an impact velocity of about 0.25 m/s and decreases continuously for increasing velocities. 
The data are fitted with the model functions for the coefficient of restitution (eq. \eqref{restmodel}) and for the fragmentation strength (eq. \eqref{fragmodel}) as described in detail in the next section. We also fitted other restitution functions for $\epsilon$, e.g. by \citet{Higa1996,Higa1998}, \citet{Andrews1930}, \citet{Borderies1984}, \citet{Thornton1998}, \citet{krijt2013}, and a number of other models taken from \citet{Antonyuk2010}. For none did we get as well matching fits. In fact, most of these models were originally used to describe restitution behavior in the inelastic regime and do not treat cohesion effects for velocities near the sticking limit. For adhesive models, fig. \ref{cmod} shows the fitted curves for the coefficient of restitution.

For velocities around the sticking-bouncing transition, all of the shown models describe the behavior of the coefficient of restitution well. For high velocities in the fragmentation regime, \citet{Thornton1998} and \cite{krijt2013} predict a slope of -1/4, which is too flat. Compared to this, the empirical function from \citet{Higa1996,Higa1998} decreases too fast. The model which we present in this work contains the observed cohesive behavior at low velocities and strikes a balance between the other models shown in the fragmentation regime. 

\begin{figure}[h]
	\includegraphics[width=\columnwidth]{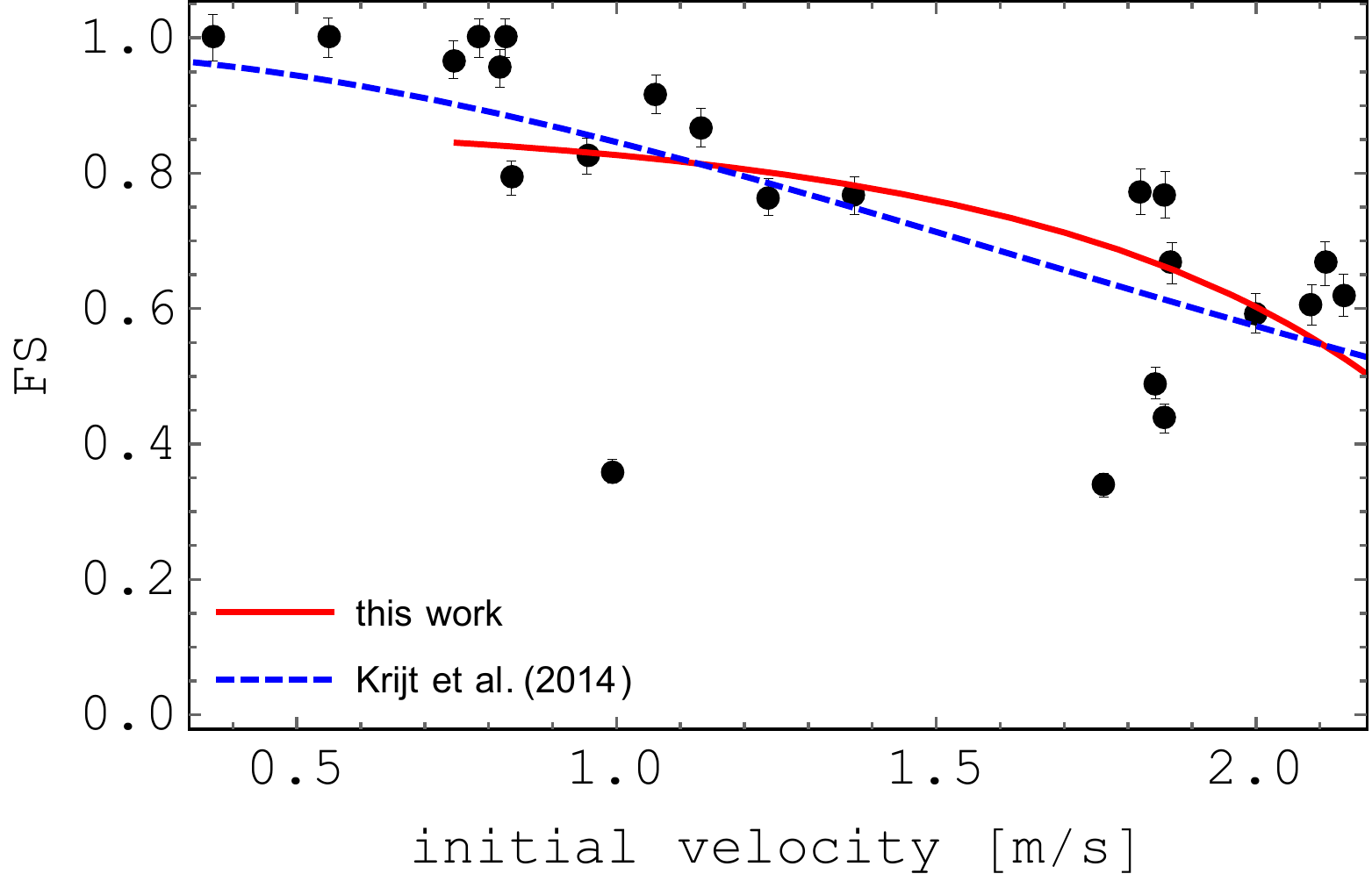}
	\caption{\label{fcom}Comparison between the fragmentation model in this work and \citet{krijt2014}.
	}
\end{figure}

Regarding the fragmentation strength, \cite{krijt2014} investigated a fragmentation model recently. Their work mainly deals with the smallest fragment after collisions. Nevertheless, they give an expression for the largest spherical fragment from which we can derive a fragmentation strength, like fig. \ref{fcom} shows. 
The curve fitted from \citet{krijt2014} fits as well to the data for the fragmentation events. However, it does not have a threshold for a transition between bouncing and fragmenting, which is clearly visible in the data. Beyond this model, we are not aware of any other analytical models for the fragmentation strength depending on the impact velocity, which we could compare to eq. \eqref{fragmodel}.

We do see sticking at low collision velocities and a transition to bouncing. However, under gravity, sticking at very low speeds and bouncing with a very small coefficient of restitution cannot be disentangled. This spatial and temporal resolution limit is reached somewhat below 5 cm/s collision velocity. Therefore, to specify a single sticking velocity $v_\text{stick}$, we do not consider the sticking collisions but use our model fit function for the coefficient of restitution. We determine the velocity for $\epsilon = 0$ and find $v_\text{stick} = (0.04 \pm 0.02)$ m/s. This sticking threshold is valid for particles with radii below 150 $\mu$m only. As larger particles do also have larger velocities, no statement on their sticking velocity can be made, as fig. \ref{CORsize} shows. 
According to \citet{Dominik1997} the sticking velocity can be calculated
as
\begin{equation}
v_\text{stick} =  \frac{1.07}{\rho^{1/2}E_\text{py}^{1/3}} \cdot   \frac{1}{R^{5/6}} \cdot  \gamma^{5/6} 
\label{stickvel}
\end{equation}
with the surface energy $\gamma$, the reduced radius $R$ and the particle mass density $\rho$.
The elastic properties enter via $E_\text{py}$, defined as $E_\text{py}=E_\text{y} / (2(1-\nu^2_\text{p}))$.
Here, $E_\text{y}$ is the Young's modulus and $\nu_\text{p}$ is the Poisson's ratio.

The first term (density, elasticity) is known for solid CO$_2$. The second term shows an almost inversely linear dependence on particle size. 
Even without knowing the third term (surface energy), the size dependence allows a comparison
of our $\rm CO_2$ data to data on silicate collisions and water-ice collisions. 
Experiments by \citet{Poppe2000} and \citet{Gundlach2015} determined the sticking velocity of 
$\sim 1 \,\rm \mu m$ grains in collisions with walls. \citet{Poppe2000} found $v_\text{stick} \approx 1.2 \,\rm m/s$ for silica spheres with a $0.6 \,\rm \mu m$ radius colliding with a wall . 
For $1.5 \,\rm \mu m$ radius water ice spheres \citet{Gundlach2015} found $v_\text{stick} \approx 9.6 \,\rm m/s$. They found roughly a factor of 10 increase in sticking velocity compared to silicates of similar size.
Using the size dependence in eq. \ref{stickvel}, we get $v_\text{stick} \approx 1.2 \,\rm m/s$  for
$1.5 \,\rm \mu m $ radius CO$_2$ grains. This does not match the values for water ice but is rather close to the value for silica of a similar grain size. This is especially true considering that the CO$_2$ grains are not spherical as there seems to be a tendency that irregular grains have somewhat higher sticking velocities in comparison to spheres \citep{Poppe2000}.

To allow an analytical treatment of CO$_2$ collisions for different parameters in the future, we extract the surface energy from our experiments as the single most important parameter. Using the longitudinal and transversal velocities $v_\text{lg}= 2900$ m/s and $v_\text{ts}=1650$ m/s of sound in solid carbon dioxide for 80 K \citep{Yamashita1997}, we can calculate Young's modulus $E_\text{y}=v_\text{lg}^2\rho \approx 13.12 \text{ GPa} $ and the bulk modulus $G = v_\text{ts}^2\rho \approx 4.25 \text{ GPa} $. Finally, the Poisson ratio is given by $\nu_\text{p}=E/(2G)-1 \approx 0.544$.  With an average particle radius of 60 $\mu$m (only particles below 150 microns are relevant for sticking), this gives a surface energy of $\gamma = 0.17^{+0.26}_{-0.13}$ J/$\text{m}^2$. The error in $\gamma$ results from the fit of the sticking velocity and the standard deviation from the size distribution. From the fragmentation model of \citet{krijt2014} we also get the surface energy as a fit parameter to $\gamma_\text{K}= (0.325 \pm 0.088)$J/m$^2$, which is consistent
with the value based on the sticking velocity. Compared to this, \citet{wood1999} estimates an surface energy for CO$_2$ between $0.08-0.097$ J/m$^2$. Given the uncertainties of the particle contacts, one should keep in mind that the slightly larger value in our experiment are effective values for the specific given setting. Within these limits the results are also consistent with the value by \citet{wood1999}.

At $v_\text{frag}= (0.75 \pm 0.05)$ m/s we find the onset of fragmentation as seen in fig. \ref{camera}.
The average kinetic energy at the onset of fragmentation is
\begin{equation}
E_\text{frag} =  \frac{1}{2}\left(\frac{4\pi\rho r_{\text{m}}^3}{3}\right)v^2_\text{frag} = (5.25 \pm 0.7)  \cdot 10^{-10} \, \text{J} \, .
\end{equation}
According to \citet{Dominik1997} fragmentation for an aggregate occurs at an energy of 
\begin{equation}
E_\text{frag} = 3 n E_\text{br}
\end{equation}
with $E_\text{br}$ as energy to break contact that is 4.5 times the sticking energy $E_\text{stick}$, or 
\begin{equation}
E_\text{frag} / E_\text{stick} = 13.5 \cdot n.
\end{equation}
In our case, it is 
\begin{equation}
E_\text{stick} = (1.38 \pm 1)\cdot 10^{-12} \,\text{J} 
\end{equation}
and we get 
\begin{equation}
E_\text{frag}/E_\text{stick} = 351
\end{equation}
for the onset of CO$_2$-aggregate fragmentation. This implies a number of contacts of 26, which is 
 consistent with the fragment count of catastrophic collisions that we observed at higher impact speeds. In detail, the fragmentation energy depends on the size of an individual aggregate in a collision and its number of contacts. We cannot quantify this in a collision at the onset of fragmentation which is not completely catastrophic. 
Therefore the considerations given before are more of a qualitative support of the aggregate nature
of larger particles. 

\begin{figure}[h]
	\includegraphics[width=\columnwidth]{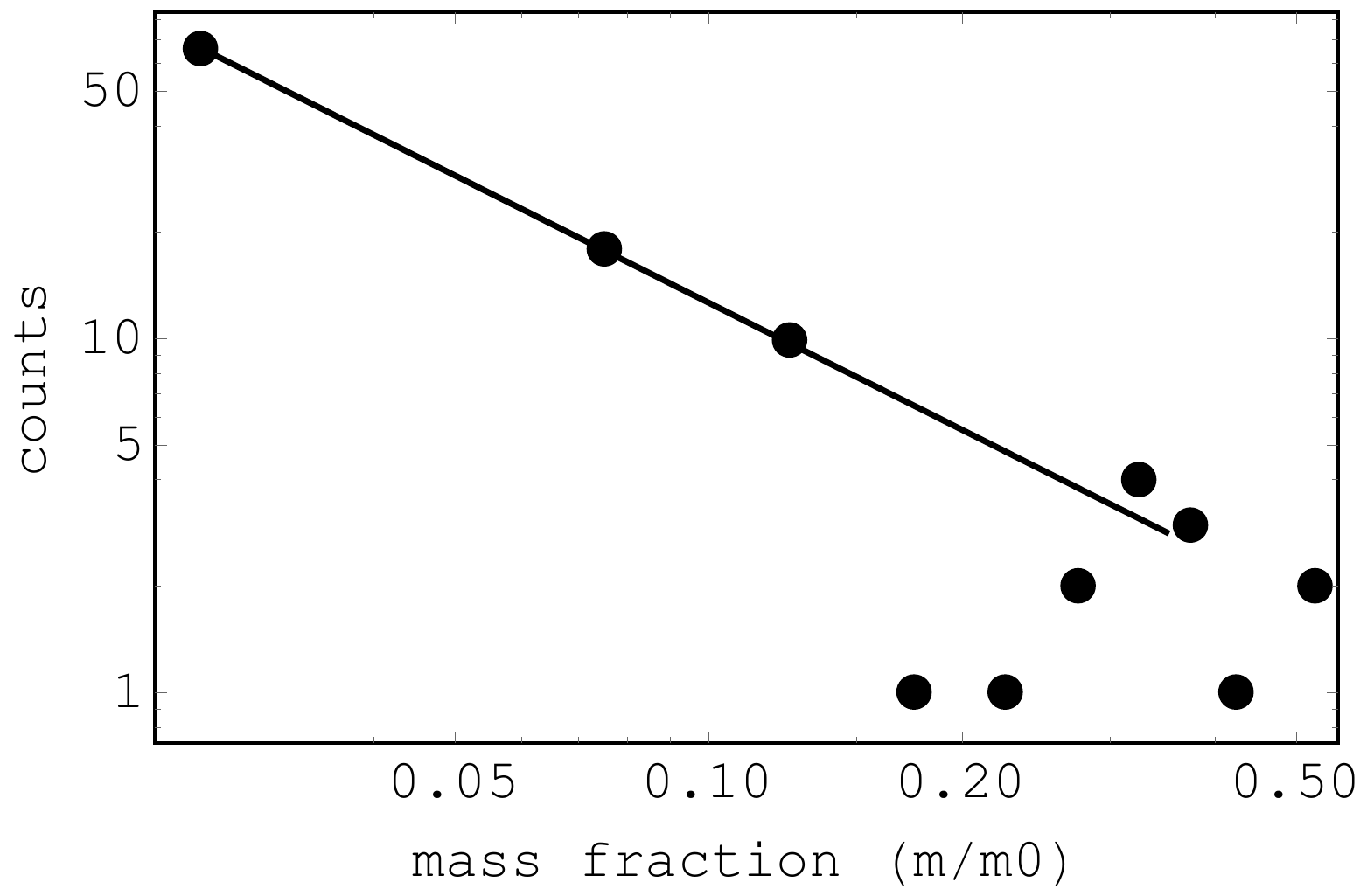}
	\caption{\label{fragmentsize} Number mass distribution of fragments after a catastrophic 
	fragmentation. $m0$ is the mass of a whole aggregate before its fragmentation.}
\end{figure}

For very large aggregates, we analyzed the mass distribution $n(m)$ of the fragments of 12 collisions, where $n(m)dm$ means that $n$ fragments are found in the mass range $[m; m+dm]$. We did not count the largest fragment and binned the data into 10 masses. The range of impact velocity varied between 0.6 and 2.2 m/s. The mass distribution is shown in fig. \ref{fragmentsize}.
This mass density function follows a power law
\begin{equation}
n \propto m^{-1.2 \pm 0.1}.
\label{massdistribution}
\end{equation}
This falls into the range found for mass density distributions for silicate aggregate collisions where power laws with exponents between -1.07 and -1.36 are found \citep{Guettler2010, Deckers2014}.
In this context, another way to characterize fragmentation is to estimate the critical fragmentation energy ($\mu = 0.5$). With $v_\text{crit}= (2.19 \pm 0.05)$ m/s we get
\begin{equation}
E_\text{crit} = (4.47 \pm 0.2) \cdot 10^{-9}  \, \text{J} \, .
\end{equation}

\section{Two new models for fragmentation and restitution}

In this section, we develop our own models for the velocity dependence of fragmentation strength and coefficient of restitution. 

\subsection{Fragmentation model}

Our fragmentation model is based on an energy balance. The energy of the aggregate with mass $M$ impacting with velocity $v_i$ has to balance the kinetic energy of all fragments after the collision $T_\text{f}$, the energy to break the contacts and account for dissipated energy  $\chi$. 
We assume that each of the $N$ fragments (subaggregates) stuck to the largest fragment with an energy of $\mathcal{V}$. 
The energy balance then reads
\begin{equation}
\frac{M}{2}v_\text{i}^2=N\mathcal{V}\Theta(v_\text{i}-v_\text{frag}) + T_\text{f} + \chi
\label{Ebal}
\end{equation}
Here, $v_\text{frag}$ is the threshold velocity for fragmentation. The Heaviside function $\Theta$ implicates that disintegrations do not happen for $v_\text{i}<v_\text{frag}$.  $T_\text{f}$ is the total kinetic energy of all fragments after the collision. We assume that $\chi$ does not depend on $v_\text{i}$ for $v_\text{i} > v_\text{frag}$. 
As seen before, the size distribution of the fragments follows a power law $n\propto a\cdot V^b$. Here, $V$ is the fragment volume, and the constants $a$ and $b$ depend on the material properties \citep{Guettler2010,Geretshauser2011}.
Therefore, we can express the fragment number in terms of volume and get
\begin{equation}
\frac{M}{2}v_\text{i}^2=\mathcal{V}\Theta(v_\text{i}-v_\text{frag})\int_{V_\text{sf}}^{V_\text{frag}}dV \, a \,V^b  + T_\text{f} + \chi \, .
\end{equation}
with $V_\text{agg} = V_\text{l}+V_\text{frag}$, where $V_\text{agg}$ is the volume of the total aggregate, $V_\text{l}$ is the volume of the largest fragment, $V_\text{sf}$ the volume of the smallest fragment, and $V_\text{frag}$ is the volume of all fragments but without the largest.
Integrated and solved for $V_\text{frag}$, this yields
\begin{multline}
V_\text{frag} = \left(\frac{(b+1)\left( \frac{M}{2}v_\text{i}^2 - T_\text{f} - \chi\right) }{a \mathcal{V}} +V_\text{sf}^{b+1}\right)^{1/(b+1)} \\
\times \Theta(v_\text{i}-v_\text{frag}).
\end{multline}
for $b<-1$.

The fragmentation strength $\mu(v_\text{i})$ can then be described by
\begin{multline}
\mu(v_\text{i}) :=  \frac{V_\text{l}}{V_\text{agg}} = \frac{ V_\text{agg}-V_\text{frag}}{V_\text{agg}}\\
=  1- \frac{1}{V_\text{agg}}\left(\frac{(b+1)\left(\frac{M}{2}v_\text{i}^2-T_\text{f}-\chi\right)}{a\mathcal{V}}+V_\text{sf}^{b+1}\right)^{1/(b+1)} \\ 
\times \Theta(v_\text{i}-v_\text{frag}),
\end{multline}
Substituting $c_1 = M(b+1)/\left(2a\mathcal{V}V_\text{agg}^{b+1}\right)$ and $c_2 = \left( (b+1)(T_\text{f}+\chi)/a\mathcal{V}-V_\text{sf}^{b+1}\right)/V_\text{agg}^{b+1}$, this simplifies to
\begin{equation}
\mu(v_\text{i}) =  1- \left[c_1 \, v_\text{i}^2 - c_2\right]^{1/(b+1)}\Theta(v_\text{i}-v_\text{frag})
\label{fragmodel}
\end{equation}
where we set $b = -1.2$ according to the measured mass distribution and set the fragmentation threshold to $v_\text{frag} = 0.75$ m/s as determined from the measurements.
Fitting this to our data results in $c_1 = (-0.073 \pm 0.027)$ s$^2$/m$^2$ and $c_2=-1.49 \pm 0.1$, which gives our model function plotted in fig. \ref{transitionplot}.

For the limiting case $T_\text{f}+\chi \gg \mathcal{V}$, we furthermore get a proportionality of the ratio between the volume of the smallest fragment and the whole aggregate and the contact energy $\mathcal{V}$ to
\begin{equation}
\frac{V_\text{sf}}{V_\text{agg}} \propto \mathcal{V}^{-1/(b+1)}.
\label{VE}
\end{equation}
	
We note that other authors like \citet{Beitz2011} or \citet{Deckers2015} describe the fragmentation strength of (dust) aggregates by a simple power law. Certainly our data could also be fitted by a power law. However, there is no physical meaning in this power law but to provide an analytic equation and the spread of data for all cited data and ours is so large that there is no clear discriminator between these functions. A striking difference is that due to the Heaviside function our model has a limited upper value for fragmentation strength somewhat below 1 if fragmentation occurs. 

\citet{krijt2013} used similar assumptions to derive the smallest size of the fragment, namely, a power law size distribution, and the dissipation of energy by creating fragments though their approach is based on surface energy of the new particle surface created. If the smallest size would be fixed e.g. due to monomers in an aggregate and the sticking energy would be size independent (s. factor $N\mathcal{V}$ from eq. \eqref{Ebal}), this should result in the same fragmentation strength model.

\subsection{Extension of the fragmentation model}

The simple fragmentation model has a cutoff in fragmentation strength somewhat below 1 at $v_\text{frag}$. We consider this sufficient for our data set and other data sets on first order (e.g.
also fig. 10 in \citet{Deckers2014} shows a cutoff). However, it does not treat the transition. In this section, we therefore extend this model slightly. We consider the dissipation $\chi(v_\text{i})$ to depend on the initial velocity $v_\text{i}$.

We mainly attribute the dissipation within the aggregate to phononic interaction. To approximate this interaction, we relate it to the harmonic oscillator so that the phonons can take the discrete energy levels $E_\text{n} = \hbar\omega (n+1/2)$ only. Furthermore, the dispersion relation should be linear $\omega \propto p$ with the momentum $p$ of the phonons. With these two assumptions, the momentum transfer $\Delta p$ to phonons will only be possible at the multiple $n$ of a discrete velocity $v_\text{R}$,
\begin{equation}
\Delta p(v_\text{i})\propto\sum\limits_{n}\delta(v_\text{i}-nv_\text{R}).
\end{equation}
To describe the macroscopic body, we smooth the delta distributions $\delta(i_\text{i}-nv_\text{R})$ with gaussian curves 
\begin{equation}
\sum\limits_{n}\delta(v_\text{i}-nv_\text{R}) \longrightarrow \sum\limits_{n} \exp\left(-(v_\text{i}-nv_\text{R})^2\right).
\end{equation}
and look at the limiting case $v_\text{R}\rightarrow0$. This provides the dissipative term $\chi(v_\text{i})$ as
\begin{multline}
\chi(v_\text{i})\propto \lim\limits_{v_\text{R}\rightarrow 0}\int\limits_{0}^{v_\text{i}}\Delta p(v_\text{i}^*)dv_\text{i}^* \\ \propto\int\limits_{0}^{v_\text{i}}\exp\left(-v_\text{i}^{*2}\right) dv_\text{i}^*\propto\text{erf}(v_\text{i})
\end{multline}
or with a constant $\chi_0$
\begin{equation}
\chi(v_\text{i})=\chi_0\text{erf}(v_\text{i}).
\end{equation}

\begin{figure}[h]
	\includegraphics[width=\columnwidth]{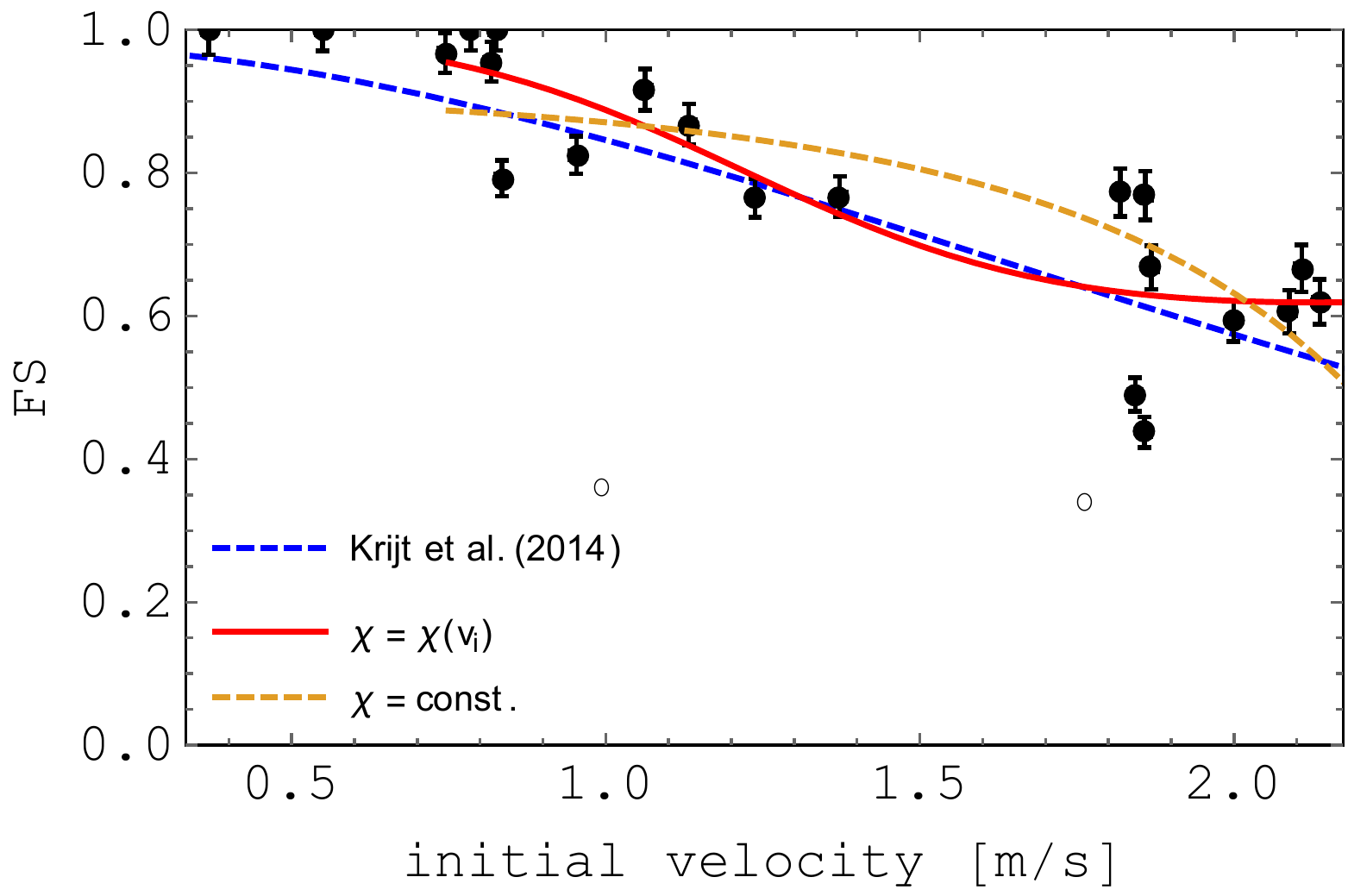}
	\caption{\label{fragX}Comparison of fragmentation models for phononic and non-phononic dissipation. The circled datapoints were not used for the fit.}
\end{figure}

For the fragmentation strength, this extension leads to
\begin{equation}
\mu(v_\text{i})=1-\left[\zeta_1v_\text{i}^2-\zeta_2\text{erf}(v_\text{i})-\zeta_3\right]^{1/(b+1)}\Theta(v_\text{i}-v_\text{frag})
\end{equation} 
with the constants $\zeta_1=c_1= M(b+1)/\left(2a\mathcal{V}V_\text{agg}^{b+1}\right)$, $\zeta_2=\left((b+1)\chi_0/a\mathcal{V}-V_\text{sf}^{b+1}\right)/V_\text{agg}^{b+1}$, and $\zeta_3= \left( (b+1)T_\text{f}/a\mathcal{V}-V_\text{sf}^{b+1}\right)/V_\text{agg}^{b+1}$. Fig. \ref{fragX} shows a comparision of the models for phononic and non-phononic dissipation.

For the constants from the fit, we get $\zeta_1 = (0.01 \pm 0.036)$ s$^2$/m$^2$, $\zeta_2= 2.32 \pm 1.01$ and $\zeta_3 = -3.5 \pm 0.9$. The benefit of the model with phononic dissipation is the behavior for $v_\text{i}\rightarrow 0$. Here, we get $\mu \rightarrow 1$ for $v_\text{frag} \rightarrow 0$.
We note that this model is capable of handling the steep transitions in \citet{Deckers2014}.
It also agrees well with better resolved transition data (Deckers and Teiser personal communication).

For a simple consideration of the fragmentation strength without smooth transitions at $v_\text{frag}$, the dependence on velocity can be neglected in the extended addend $\zeta_2\text{erf}(v_\text{i})$. As a result, the data can then be fitted with two parameters only, like shown in fig. \ref{transitionplot}.

\subsection{Restitution model}

\begin{figure}[h]
	\includegraphics[width=\columnwidth]{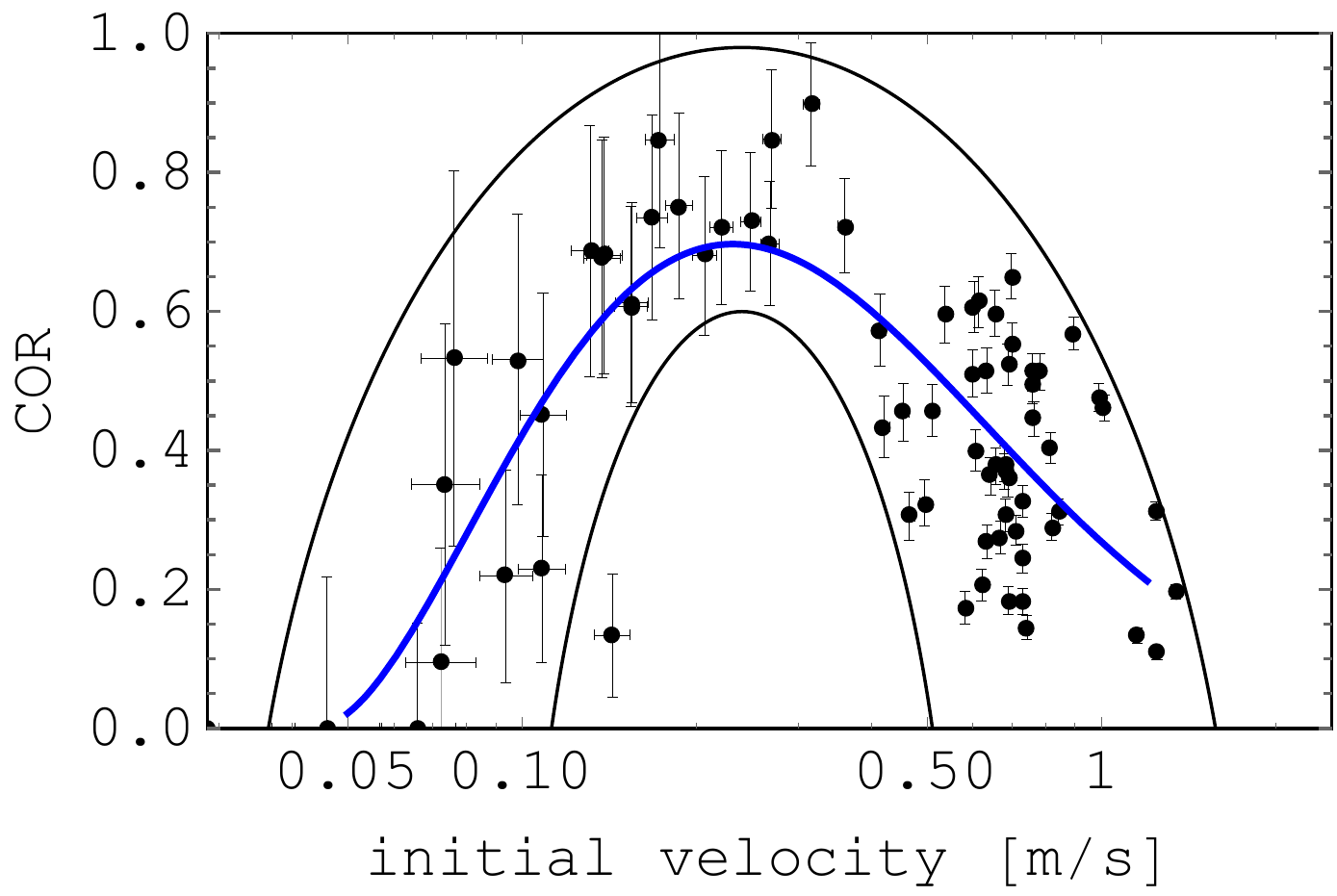}
	\caption{\label{rm} Coefficient of restitution on a logarithmic scale. An elliptical envelope is overplotted around the axis of symmetry.}
\end{figure}

As a starting point, we take a look at the coefficient of restitution on a logarithmic scale, as fig. \ref{rm} shows.
Overplotting an elliptical envelope centered at a velocity $v_\text{c}$ suggests that the shape of the data implies the existence of a line symmetry as soon as we transform $v_\text{i} \rightarrow \tilde{v_\text{i}} = \ln\left(v_\text{i}/v_\text{c}\right)$. Forces, which are responsible for the cohesion, seem to have a similar nature to forces between contacts of grains, which have to be overcome for plastic deformation on a multiplicative scale.

The enclosing envelope $G_\epsilon(v_\text{i})$ can be parameterized by
\begin{equation}
  G_\epsilon(v_\text{i})= \delta_1 + \sqrt{\delta_2 + \delta_3 \ln\left(\frac{v_\text{i}}{\delta_4}\right)^2}
\end{equation}
with $\delta_1=-0.5, \delta_2=2.19, \delta_3=-0.55, \delta_4=0.24$ m/s for the upper boundary and $\delta_1=-0.5, \delta_2=1.2, \delta_3=-1.7, \delta_4=0.24$ m/s for the lower boundary. $\delta_4$ describes the critical velocity $v_c := \delta_4$.
	
Once this symmetry applies, the derivative $\epsilon'(\tilde{v_\text{i}})$ has to be point symmetrical in $\tilde{v_\text{i}}$. Here, we approximate this by a linear symmetry $\epsilon'(\tilde{v_\text{i}})/\epsilon(\tilde{v_\text{i}})= a_1\tilde{v_\text{i}}$ with a constant $a_1$. This might also be written as 
\begin{equation}
\frac{\epsilon'(\tilde{v_\text{i}})}{\epsilon(\tilde{v_\text{i}})}=\frac{d}{d\tilde{v_\text{i}}}\ln\left(\epsilon(\tilde{v_\text{i}})\right)= a_1\tilde{v_\text{i}}.
\end{equation} 
Integrating and solving for $\epsilon(\tilde{v_\text{i}})$, this gives
\begin{equation}
\epsilon(\tilde{v_\text{i}}) = \exp\left(a_1\tilde{v_\text{i}}^2+a_2\right)
\end{equation} 
with a second constant $a_2$. Now we can transform back to a term depending on $v_\text{i}$. This finally leads to
\begin{equation}
\epsilon(v_\text{i}) = A\cdot\exp\left(a_1\left(\ln\left(\frac{v_\text{i}}{v_\text{c}}\right)\right)^2\right)
\label{restmodel2}
\end{equation} 
with $A=\exp(a_2)$. 

In the experiment we point out a sticking velocity $v_\text{stick}$, which eq. \eqref{restmodel2} does not consider so far. We can fulfill this condition with a linear transformation $v_\text{i} \rightarrow v_\text{i} - v_\text{stick}$ and multiplication by $\Theta(v_\text{i}-v_\text{stick})$, so that $\epsilon(v_\text{i})=0 \hspace{3mm} \forall \hspace{3mm} v_\text{i}<v_\text{stick}$. In total, we gain a function for the coefficient of restitution
\begin{equation}
\epsilon(v_\text{i}) = A\cdot\exp\left(a_1\left(\ln\left(\frac{v_\text{i}-v_\text{stick}}{v_\text{c}}\right)\right)^2 \right)\Theta(v_\text{i}-v_\text{stick})
\label{restmodel}
\end{equation} 
Strictly speaking, this transformation breaks the assumed symmetry to some extent. The coefficient of restitution decreases more in the neighborhood of the sticking velocity than for high velocities, which can be observed in fig. \ref{rm}.

For our data, we get $A = 0.67 \pm 0.04, v_\text{c} = (0.189 \pm 0.025)$ m/s, $a_1= -0.36 \pm 0.04$, and $v_\text{stick} = (0.04 \pm 0.02)$ m/s. The dependency in eq. \eqref{restmodel} has a universal nature. For $v_\text{stick}=0$, it gives the lognormal-distribution, which is used widely, for example, in fitting distributions of particle sizes \citep{Raabe1971}, describing the magnetization behavior of nm-Al-particles \citep{Buhrman1976} or even biological systems \citep{Koch1966}. For $a_1=-1$ and $v_\text{stick}=0$, we obtain the empirical equation from \citet{Higa1996, Higa1998}, where 
\begin{equation}
\epsilon(v_\text{i})=A\cdot\left(\frac{v_\text{i}}{v_\text{c}}\right)^{-\ln\left(\frac{v_\text{i}}{v_\text{c}}\right)}
=A\cdot\exp\left(-\left(\ln\left(\frac{v_\text{i}}{v_\text{c}}\right)\right)^2\right)
\label{higaeq}
\end{equation}
was used to describe the collision behavior of centimeter-sized H$_2$O-ice spheres in the inelastic region.

\section{Consequences for planet and comet formation}

If grains outside the CO$_2$ ice line have onion like shells, the outer CO$_2$ shell is thick
enough to dominate the sticking properties. Certainly, fragmentation of shell-like bodies 
might be different \citep{Geretshauser2011}. However, for the low-energy collisions considered here,
we regard the pure CO$_2$ collisions as well-suited analog. We picture a possible simple result
of our studies in fig. \ref{ppde}.

If no further recycling and phase transitions occur, there will be a torus of several AU (zone 2)
between water ice line and CO$_2$ ice line where planetesimals form preferentially. Farther in and farther out, growth will proceed slower or get stuck, e.g. at bouncing barriers before instabilities can kick in. Another possibility is that larger ice structures could form in upper (warmer) disk layers, where H$_2$O ice in contrast to CO$_2$ ice is still present. Those aggregates could cross the CO$_2$ snow line due to vertical settling resulting in sweep up of smaller CO$_2$ coated particles and finally runaway growth of a limited number of aggregates as, e.g. described for "lucky winners" in \citet{Windmark2012}.

Other scenarios are possible as well. Gravitational instability, for example, can be partly sensitive to
certain sizes. Therefore, it might be beneficial if growth does not proceed to large structures. In this case, zone 1 or zone 3 might be favored for planetesimal formation.
Also, in the vicinity of the ice lines and especially in turbulent disks phase transitions might be
frequent with sublimation and recondensation changing the structure of the solids constantly.
This might also be beneficial to form larger bodies \citep{Ros2013}. Nevertheless, CO$_2$ collisions would dominate the further evolution also in this picture at later times. 

This shows some potential outcomes, which
require more complex modeling, but put emphasize on the importance of CO$_2$ collisions.
It is crucial to note that ices can behave very differently. We only consider CO$_2$ ice in this work but 
the concept of ice lines does not end here. CO was already mentioned above. How CO collisions 
proceed is unknown. Considering the dipole moment as a crucial factor for the collisional outcome, CO should behave similar to CO$_2$ ice since the dipole moment of CO of 0.112 D is very small compared to the dipole moment of H$_2$O of 1.84 D \citep{Scuseria1991}. There are also "high temperature" ice lines. \citet{Lodders2004} e.g. propose the 
existence of a tar line inside of the water ice line to explain compositional details of Jupiter.

The similarity to silicate properties does not necessarily prevent formation of planetesimals in zone 3 as there are terrestrial planets supposed to be formed in zone 1. However, formation timescales and final size and composition of the forming objects (comet, planetesimal, or planet) might be different.

\section{Conclusion}
We measured the threshold impact speeds between sticking and bouncing and between bouncing and fragmentation of $\sim$ 100 $\mu$m sized CO$_2$-ice aggregates. We obtain a sticking threshold of $v_\text{stick} = (0.04 \pm 0.02)$ m/s. Scaled in size, this is similar to silicates \citep{Poppe2000} but a factor of 10 less than for water ice \citep{Gundlach2015}. We derive a surface energy of $\gamma = 0.17^{+0.26}_{-0.13}$ J/$\text{m}^2$ for CO$_2$ ice.
After the bouncing regime fragmentation sets in at $v_\text{frag}= (0.75 \pm 0.05)$ m/s. Critical fragmentation occurs at  $v_\text{crit}= (2.19 \pm 0.05)$ m/s. 
The size distribution of fragments except the largest one follows a power law $n \propto m^{-1.2 \pm 0.1}$, consistent with aggregates of silicate grains \citep{Guettler2010}. 
The critical fragmentation energy is also comparable to the energy deduced from tensile strength measurements of 100 $\mu$m basalt-, and palagonite aggregates \citep{debeule2015}.
This all implies that CO$_2$ ice behaves like silicates in collisions.

We further developed a model for fragmentation (eq. \eqref{fragmodel}) and restitution (eq. \eqref{restmodel}). Within this model, the fragmentation strength depends mainly on the size distribution of the fragments, which has to be known from empirical measurements. We obtain a similar equation for the coefficient of restitution as \citet{Higa1996, Higa1998} used for centimeter-sized H$_2$O-ice spheres.

To conclude, the term ice in planetesimal formation is often used with the tacit understanding
that it is water ice. Water ice is more sticky due to the dipole interactions. It is
tempting to assume that beyond the water snow line the stickiness increases for all kinds of particles and aggregates.
As shown  here, this is not the case. Collisions of (pure) CO$_2$ ice grains are much more comparable
to the collisions of silicate grains. If CO$_2$ covers water-ice grains beyond 10 AU, 
collisions farther out will be less sticky then farther inward. This restricts a 
water-ice-dominated preferential planetesimal formation to a torous of a few AU between
the water ice line and the CO$_2$ ice line.

\section{Acknowledgements}

This work is supported by the DFG under the grant number WU321/12-1 and TE890/1-1.

\bibliographystyle{apj}
\bibliography{references}

\end{document}